\documentclass[preprint]{elsarticle}

\usepackage{lineno,hyperref}
\modulolinenumbers[100]

\journal{Physics of the Dark Universe}

\begin{document}

\begin{frontmatter}

\title{Perturbative Resonance in WIMP paradigm and its Cosmological Implications on Cosmic Reheating and Primordial Gravitational Wave Detection}

\author{Changhong Li\corref{mycorrespondingauthor}}
\address{Department of Astronomy, Key Laboratory of Astroparticle Physics of Yunnan Province,  \\ School of Physics and Astronomy,  Yunnan University, \\ No.2 Cuihu North Road, Kunming,  650091 China}
\cortext[mycorrespondingauthor]{Corresponding author}
\ead{changhongli@ynu.edu.cn}

\begin{abstract}
We investigate the co-evolution of dark matter (DM) density perturbation and metric perturbation in the WIMP paradigm. Instead of adopting the conventional assumption that DM starts out in thermal equilibrium, we propose a simple phase of DM production for the WIMP paradigm and extend our analysis to this phase. Being free from the envelop of thermal equilibrium, an amplified perturbative resonance between DM density perturbation and scalar modes of metric perturbation takes place during the DM production phase, and consequently results in a suppression of the tensor-to-scalar ratio of metric perturbation. By specifying the cosmic background with a typical realization of cosmic reheating, we establish a  relation between DM particle mass $m_\chi$ and the tensor-to-scalar ratio $r$ in the WIMP paradigm, which also contains two reheating parameters, the reheating temperature $T_{R_f}$ and the dissipative constant $\Gamma_0$. Notably, for a sizeable parameter region of WIMP candidate and cosmic reheating, this relation predicts a smaller value of $r$ in comparing with the conventional expectation obtained by assuming DM starts out in thermal equilibrium. Once the suppression of $r$ is measured in future observations of primordial gravitational wave in CMB,  this relation can be used to constrain $m_\chi$, $T_{R_f}$ and $\Gamma_0$ in principle. 
\end{abstract}


\end{frontmatter}

\linenumbers

\section{Introduction}
\label{sec:intro}
The particle nature of dark matter (DM) is one of the most important topics in modern physics. But current evidence is  insufficient to settle down such microscopic issue -- as existing evidence primarily reflects the macroscopic properties of DM fluid/cloud/halo \cite{Bertone:2016nfn}. Nowadays, most compelling attempts for unveiling DM particle nature mainly rely on the (in)-direct detections or reproduction of DM particles \cite{Tanabashi:2018oca, Conrad:2017pms, Aaboud:2017phn}. Unfortunately, due to its tiny cross-section, DM particle may not be detected soon \cite{Bertone:2018xtm}. In particular, for the weakly interacting massive particle (WIMP), the well-known logarithmic relation between DM particle mass $m_\chi$ and its thermally averaged cross-section $\widetilde{\langle\sigma v\rangle}$ has been established by using the background DM density fraction $\Omega_\chi=0.26$ for a long time. However, neither of them can be determined separately. To go beyond such background issue, we are motivated to study the evolution of DM density perturbation in the early Universe as it may contain more information about the nature of DM particle. 

\begin{figure}[tbp]
\centering
\includegraphics[width=.8\textwidth]{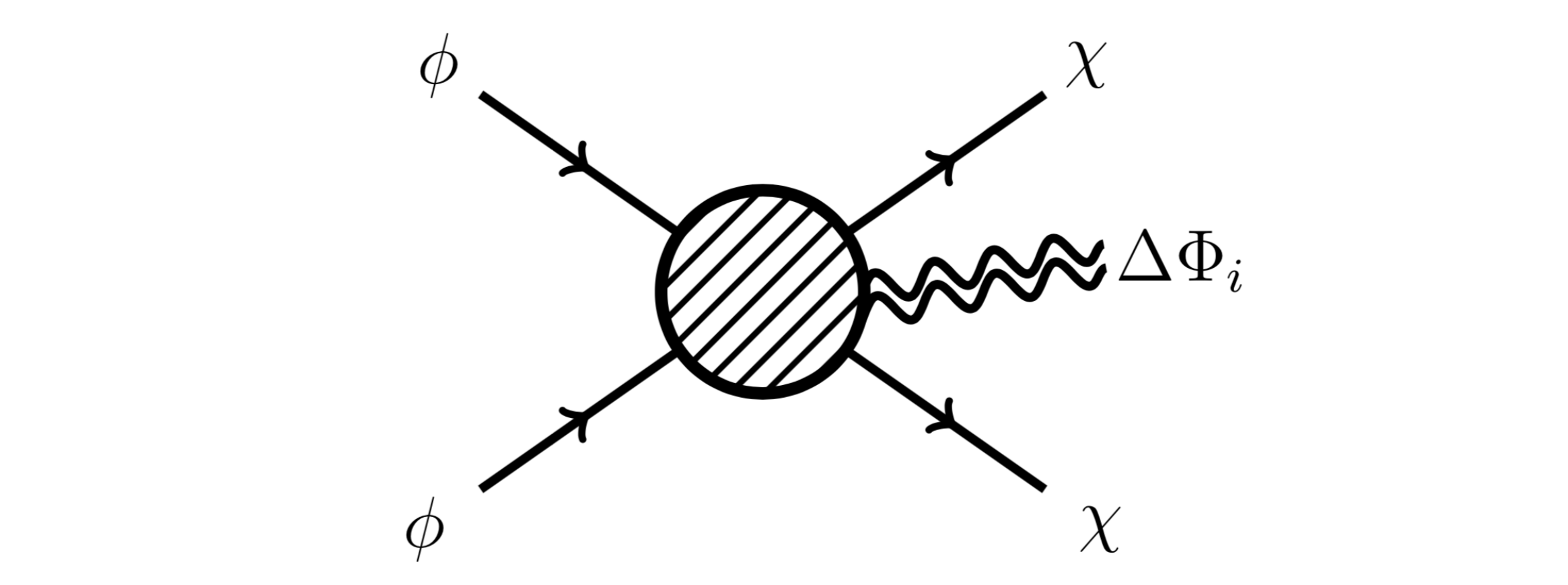}
\caption{\label{fig:feyman.pdf} An illustration of a pair of DM particles $\chi$, produced via a pair annihilation of scalar particles $\phi$ in a curved and nonrigid spacetime ({\it c.f.} FIG. 1(b) in \cite{Li:2019cjp}). Driven by the chemical potential difference between $\chi$ and $\phi$, each pair production of $\chi$ can generate a small local scalar fluctuation  $\Delta\Phi_i$ on spacetime. During an out-of-chemical equilibrium phase such as DM production, such metric fluctuations can accumulate and contribute to the background metric perturbation.}
\end{figure}

In a recent work \cite{Li:2019cjp}, by investigating the co-evolution of DM density perturbation and metric perturbation in the non-thermal equilibrium DM (NEQDM) model, a perturbative resonance\footnote{We use ``perturbative resonance'' to emphasize that, driven by the new terms on the right-hand side of Eq.(\ref{eq:eompert}), the metric and density perturbations are amplified simultaneously, especially in the same period (the same wave-vector $k$) following a positive feedback pattern. One should notice that such ``resonance'' is not similar to the well-known Breit-Wigner resonance.}, which is driven by the pair productions of DM particles, is unveiled. As illustrated in FIG.\ref{fig:feyman.pdf}, in a curved and non-rigid spacetime, driven by the chemical potential difference between DM particle $\chi$ and a lighter scalar particle $\phi$, each pair production of DM particles can produce a small but non-zero fluctuation in the trace of metric, $\Delta\Phi_i$. In a thermal equilibrium background, such extra metric fluctuations must be cancelled exactly by each other on the interaction-by-interaction basis. However, they can accumulate during an out-of-thermal equilibrium phase such as the DM production phase and then they will contribute to the background metric perturbation. Specifically, as shown in Ref.\cite{Li:2019cjp}, during the phase of DM production in the NEQDM model, the accumulation of these fluctuations causes an amplified perturbative resonance between DM density perturbation and scalar modes of metric perturbation, and eventually results in a suppression of the tensor-to-scalar ratio of metric perturbation $r$. By solving the Einstein-Boltzmann equation governing such perturbative resonance, a relation between $m_\chi$ and $r$ is established for the NEQDM model in that work. As such result can relate the particle nature of NEQDM candidate to the future observations of B-modes in CMB, we, in this paper, are motivated to apply a similar strategy to investigate the perturbative resonance in the WIMP paradigm.       

The previously studied NEQDM belongs to a broader class of freeze-in models of dark matter \cite{Baer:2014eja,Chung:1998ua,Shi:1998km, Feng:2003xh, Hall:2009bx, Cheung:2011nn, Klasen:2013ypa, Feldstein:2013uha, Cheung:2014nxi, Cheung:2014pea}. The primary distinction between the WIMP DM and NEQDM is that the WIMP DM can obtain the thermal equilibrium, but NEQDM (and other Freeze-in models) never obtain it. The difference implies that the perturbative resonance induced by their production would end up in two different ways. For WIMP DM, the perturbative resonance ends when WIMP DM attains thermal equilibrium. Nevertheless, for NEQDM, the perturbative resonance ends after the background temperature falls below its particle mass.
Consequently, the analysis and results of these two models are quite different, as we will show in the following sections~\footnote{For example, there are three phases in WIMP to be analyzed (Production, Equilibrium, and Freeze-out ). However, in NEQDM, there are only two phases (Production and Freeze-in).}. Moreover, investigating such an effect for WIMP is helpful for better understanding the standard WIMP paradigm, which is an essential part of standard cosmology. 

In the standard WIMP scenario \cite{Dodelson:2003ft}, it is typically assumed that DM starts in thermal equilibrium and the production phase is instant ($\Gamma_{WIMP}\gg H$), as such assumption does not affect the relic abundance \cite{Bertone:2004pz}. However, with such oversimplified assumption, these extra metric fluctuations are not supposed to accumulate in its initial  thermal equilibrium background. Therefore, the co-evolution of DM density perturbation and the scalar modes of metric perturbation is trivial and no perturbative resonance can be observed in the conventional WIMP paradigm.    

To overcome this obstacle, we embed a simple phase of DM production into the standard WIMP paradigm. By doing so, we obtain a pre-thermal equilibrium WIMP DM model that we can study the evolution of DM density perturbation during the post-reheating DM production phase. For this pre-thermal equilibrium WIMPs scenario, it initially has a era of freeze-in-like production, but then obtains thermal equilibrium, which the NEQDM (and other Freeze-in models) never obtain. Moreover, if the production phase is instant in the pre-thermal equilibrium WIMPs, the standard WIMP is recovered~\footnote{More specifically, as we will show in Figure~\ref{fig:Phi.pdf}, for the pre-thermal equilibrium WIMP picture explored in this paper, the standard WIMP picture is recovered by taking the coupling constant $\lambda$ to be large enough.}.    

By using the same methodology borrowed from Ref.\cite{Li:2019cjp}, we demonstrate that, in the pre-thermal equilibrium WIMP scenario, driven by pair productions of DM particles, an amplified perturbative resonance between DM density perturbation and scalar modes of metric perturbation can also take place. As such resonance does not affect the tensor modes of metric perturbation, it results in a suppression of tensor-to-scalar ratio $r$ when the scalar modes are amplified during the phase of DM production. Explicitly, by solving the improved Einstein-Boltzmann equation, we find that, in presence of the perturbative resonance, the final value of $r$ is not solely determined by its primordial value $r_i$, which was obtained by assuming DM starting out in thermal equilibrium, but also relies on the DM abundance at the end of reheating $Y_{R_f}$. Thus, by using a typical reheating model \cite{Bassett:2005xm, Allahverdi:2010xz, Amin:2014eta} to specify the relation between $Y_{R_f}$ and $m_\chi$, we establish a new relation between $r$ and $m_\chi$ in the pre-thermal equilibrium WIMP scenario, which also contains two reheating parameters, the reheating temperature $T_{R_f}$ and the dissipative constant $\Gamma_0$. According to this relation, for a sizeable parameter region of WIMP candidate and cosmic reheating, the value of $r$ predicted in the pre-thermal equilibrium WIMP scenario is much smaller than the conventional expectation obtained by assuming DM starting out in thermal equilibrium. Once a smaller $r$ is measured in future observations of B-modes in CMB \cite{Abazajian:2016yjj}, it would imply that the perturbative resonance plays an important role during the production of pre-thermal equilibrium WIMPs. And this new relation would allow $m_\chi$,  $T_{R_f}$ and $\Gamma_0$ to be constrained accordingly. Notably, this relation based on perturbative resonance is not only alternative/complimentary to existing strategies based on DM abundance \cite{Patrignani:2016xqp}, but also suggests that the primordial gravitational wave can serve as a new probe for indirect DM searches, which is beyond the conventional gamma and cosmic rays \cite{Conrad:2017pms}.

\section{Cosmic background and Dark Matter abundance} \label{sec:modelling}
In this section, we discuss the evolution of DM abundance in the pre-thermal equilibrium WIMP scenario. To realize a typical process of WIMP paradigm, we consider the simplest case that DM particles $\chi$ are produced by the pair annihilations of a lighter scalar particles $\phi$ with the minimal coupling, $\mathcal{L}=\lambda \phi^2\chi^2$ ~\footnote{To avoid ruining the Big bang nucleosynthesis (BBN), $\phi$ must freeze out much earlier than $T_{BBN}\sim 1~\textbf{MeV}$, which implies $m_\phi \gg 1~\textbf{MeV}$. }. In particular, we assume that $m_\phi\ll m_\chi$, $\phi$ is tightly coupled to cosmic bath for $T\ge m_\phi$, and $\lambda$ is large enough that $\chi$ can attain thermal equilibrium before it freezes out. In this study, we also take the standard cosmological model \cite{Dodelson:2003ft} as the fiducial cosmic background. 

\begin{figure}[tbp]
\centering
\includegraphics[width=.8\textwidth]{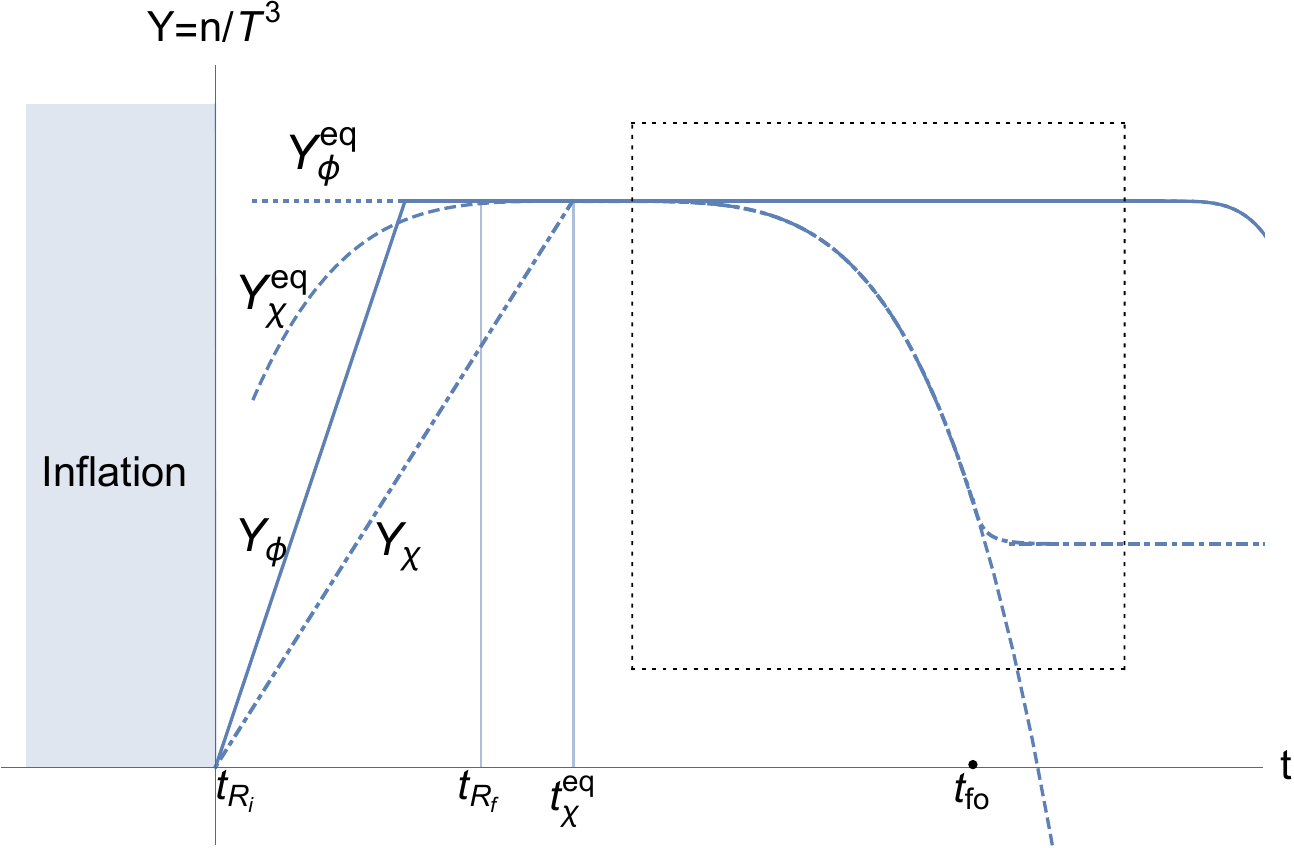}
\caption{\label{fig:capab.pdf} As schematic plot of the pre-thermal equilibrium WIMP scenario. After inflation, the Universe is reheated during $t_{R_i}\le t\le t_{R_f}$. As the lighter scalar particles $\phi$ are strongly coupled to the cosmic bath, its abundance $Y_\phi$ (the solid lines) can immediately attain its thermal equilibrium $Y_\phi^{eq}$ (the dotted line) and then tracks it. However, due to the tiny cross-section, the process of DM production goes longer into cosmic reheating. At $t=t_\chi^{eq}$, the abundance of DM particles $Y_\chi$ (the dash-dotted lines) attains its thermal equilibrium $Y_\chi^{eq}$ (the dashed curve). After that, DM abundance tracks it until freezing out at $t=t_{fo}$. Note that, the conventional WIMP paradigm, which ignores the process of DM production, is in the black dotted box. Note that $Y_\phi$ also displays standard freeze-out behavior at the late time ($m_\phi<T$), which is not plotted in this schematic plot. And the plot of the evolution of $Y_\phi$ and $Y_\chi$ before obtaining the equilibrium values are schematic for illustration.}
\end{figure}

To embed the phase of DM production to a generic WIMP paradigm, we assume that the abundances of $\chi$ and $\phi$ are zero at the end of inflation, $n_\chi(t_{R_i})=n_\phi(t_{R_i})=0$.  After $t=t_{R_i}$, the inflaton $\varphi$ decays into Standard Model (SM) particles, and reheats Universe to the highest temperature, $T=T_{R_f}$, at $t=t_{R_f}$. The interval $t_{R_i}\le t\le t_{R_f}$ refers as cosmic reheating. As illustrated in FIG.\ref{fig:capab.pdf}, the particles $\phi$ are produced efficiently from cosmic bath during the reheating and its abundance (the solid lines) attains and tracks its thermal equilibrium ($n_\phi=n_\phi^{eq}$) immediately. 

However, due to its tiny cross-section, DM particles $\chi$ are not able to attain thermal equilibrium at the end of reheating ($t=t_{R_f}$). The process of DM production can go longer into cosmic reheating and cease until attaining thermal equilibrium at $t=t_\chi^{eq}$. After that, DM abundance tracks its thermal equilibrium until freezing out at $t=t_{fo}$. As shown in FIG.\ref{fig:capab.pdf}, the conventional WIMP paradigm in the black dotted box is only a part of this pre-thermal equilibrium WIMP scenario. Except the phase of DM production, the pre-thermal equilibrium WIMP scenario is the same to the conventional WIMP paradigm.       

Analytically, by decomposing the unintegrated Boltzmann equation, $df/dt=C[f]$, in the curved spacetime of early Universe, $g_{\mu\nu} = \{-1-2\Psi(\vec{x},t), a^2(t)\delta_{ij}\left[1+2\Phi(\vec{x},t)\right]\}$, we can obtain the equations of motion for the background DM abundance and the DM density perturbation, Eq.(\ref{eq:eomback}) and  Eq.(\ref{eq:eompert}), respectively. For the background DM abundance, it is governed by
\begin{equation} \label{eq:eomback}
\frac{dn_\chi}{dt}+3Hn_\chi=\widetilde{\langle\sigma v\rangle}\left[\left(\frac{n_\phi}{n_\phi^{eq}}\right)^2\left(n_\chi^{eq}\right)^2-n_\chi^2\right]~,
\end{equation} 
where $H$ is the Hubble parameter, $\widetilde{\langle\sigma v\rangle}$ is the thermally averaged cross-section\footnote{The definition of the thermally averaged cross section for the reaction $1+2\leftrightarrow 3+4$ takes the following form, 
\begin{eqnarray}
\nonumber    &&\widetilde{\langle\sigma v\rangle}\equiv \frac{1}{n_1^{eq} n_2^{eq}}\int\frac{d^3 p_1}{(2\pi)^3 2 E_1}\int\frac{d^3 p_2}{(2\pi)^3 2 E_2}\int\frac{d^3 p_3}{(2\pi)^3 2 E_3}\int\frac{d^3 p_4}{(2\pi)^3 2 E_4} e^{-(E_1+E_2)/T}\\
 \nonumber   && \times (2\pi)^4 \delta^3(p_1+p_2-p_3-p_4)\delta(E_1+E_2-E_3-E_4)|\mathcal{M}|^2~,
\end{eqnarray}
where the amplitude $\mathcal{M}$ is determined by $\lambda$ in our model. The component $e^{-(E_1+E_2)/\mu}$ in the integral  implies that the particles are in a thermal distribution ({\bf kinetic equilibrium}). In practice, such a condition is easy to be attained as the kinetic collision in generic takes place rapidly. More specifically, in this work, we assume that, after out-of-equilibrium production, although not in {\bf thermal equilibrium}, DM (and $\phi$) can swiftly attain a thermal distribution ({\bf kinetic equilibrium}) through the efficient kinetic collisions within the cosmic bath. For other exceptions that cannot obtain kinetic equilibrium, this issue is quite model-dependent but worthy of further studies. }, $f$ is the distribution function, and the form of the collision function $C[f]$ is determined by $\mathcal{L}_{int}=\lambda\phi^2\chi^2$. In our analysis, $g_{\mu\nu}$ is taken as the perturbed Friedmann-Lemaitre-Robertson-Walker (FLRW) metric in the conformal Newtonian gauge. Note that, this equation is not only able to describe the freezing-out process but also able to describe the process of DM production ($n_\phi\le n_\phi^{eq}$ and $n_\chi\le n_\chi^{eq}$). The extra factor $\left(n_\phi/n_\phi^{eq}\right)^2$ accounts for the earliest stage of the evolution and it vanishes when $\phi$ attains its thermal equilibrium $n_\phi=n_\phi^{eq}$.  

By solving Eq.(\ref{eq:eomback}) with the initial condition $n_\chi(t_{R_i})=n_\phi(t_{R_i})=0$, we  obtain,
\begin{equation} \label{eq:ychib}
Y_\chi=\frac{1-e^{-2\pi^2[\kappa(y-y_{R_f})+Y_{R_f}]}}{\pi^2\left(1+e^{-2\pi^2[\kappa(y-y_{R_f})+Y_{R_f}]}\right)}, \qquad y_{R_f}\le y\le 1~, 
\end{equation}
where $Y_\chi\equiv n_\chi T^{-3}$ is DM abundance, $Y_{R_f}$ is its value at the end of reheating $t=t_{R_f}$, and $y\equiv m_\chi T^{-1}$ is a re-parameterized variable that relates to $t$ as $dy/dt=Hy$ during the post-reheating radiation-dominated era. The newly introduced parameter $\kappa\equiv m_\chi^3\sigma_0(4\pi^4H_m)^{-1}$ is a dimensionless constant that is defined with the reduced Hubble parameter $H_m\equiv Hy^2$ and the reduced thermally averaged cross-section $\sigma_0=4\widetilde{\langle\sigma v\rangle}y^{-2}$. In particular, to realize WIMP paradigm, DM should be able to attain thermal equilibrium before freezing-out, so that it requires $\kappa\gg (2\pi^2)^{-1}$ \cite{Li:2014era}. Moreover, we should notice that $Y_{R_f}$ can not be expressed as a function of $y$. To get its expression, we need to integrate Eq.(\ref{eq:eomback}) during the reheating,
\begin{equation}\label{eq:yrfexi}
Y_{R_f}=T_{R_f}^{-3}\int_{t_{R_i}}^{t_{R_f}}\widetilde{\langle\sigma v\rangle}\left(n_\phi/n_\phi^{eq}\right)^2\left(n_\chi^{eq}\right)^2 dt~.
\end{equation} 
This expression reflects that $Y_{R_f}$ is a crucial parameter for constraining $m_\chi$ and $T_{R_f}$ as it not only encodes the nature of DM particles  but also contains the information of cosmic reheating.  Although $Y_{R_f}$ is sub-leading in background DM abundance and will be overwhelmed in thermal equilibrium, as we will show in next section, it can be traced from the evolution of DM density perturbation. This observation implies that, by constraining the value of $Y_{R_f}$ with the co-evolution of DM density perturbation and metric perturbation, $m_\chi$ and $T_{R_f}$ can be constrained accordingly.

Now we use Eq.(\ref{eq:ychib}) to summarize the evolution of DM abundance. At first, during DM production ($t_{R_i}\le t \le t_\chi^{eq}$), its abundance is much smaller than the thermal equilibrium. By expanding Eq.(\ref{eq:ychib}) with respect to the small $\kappa y$, we can obtain $Y_\chi=\kappa(y-y_{R_f}+\kappa^{-1}Y_{R_f})$ for this phase. At $t=t_\chi^{eq}$, DM abundance attains thermal equilibrium, $Y_\chi=Y_\chi^{eq}=\pi^{-2}$,  and then tracks it until freezing out. With the falling of background temperature in the radiation-dominated expansion, the thermal decoupling takes place at $T\simeq m_\chi$ ($t=t_{fo}$). Then DM freezes out to achieve a decreased relic abundance, $Y_f=(4\pi^6\kappa)^{-1}$. For short, we list the whole evolution of DM abundance in the pre-thermal equilibrium WIMP scenario as 
\begin{equation} \label{eq:wholenwimp}
Y_\chi(y)=\left\{  
\begin{array} {l}
 {\displaystyle \kappa(y-y_{R_f}+\kappa^{-1}Y_{R_f}) ~, \qquad y_{R_f}\le y\le y_\chi^{eq}}; \qquad \textbf{Production} \\ 
 \\ 
  {\displaystyle  \pi^{-2} ~, \qquad\qquad\qquad\qquad\quad~~ y_\chi^{eq}\le y\le y_{fo}}; \qquad \textbf{Equlibrium}    \\
  \\
    {\displaystyle  (4\pi^6\kappa)^{-1} ~, \qquad\qquad\qquad\qquad\quad y\ge y_{fo}}; \qquad\quad \textbf{Freezing-out}    \\
\end{array}     
\right. ,
\end{equation}
where $y^{eq}=\left(2\pi^2\kappa\right)^{-1}$ and $y_{fo}\simeq 1$ correspond to $t_\chi^{eq}$ and $t_{fo}$ respectively. By using the third line of Eq.(\ref{eq:wholenwimp}), we have $Y_f=1.7\times 10^{-29} (\sigma_0 m_\chi)^{-1} \textbf{eV}^{-1}$. By imposing the observed density fraction of DM, $\Omega_\chi=1.18\times 10^{-2} \textbf{eV}^{-1}\times m_\chi Y_f= 0.26$, we reproduce the well-known prediction, $\sigma_0\simeq 0.3\times 10^{-39}\textbf{cm}^2$ ({\it c.f.} Eq.(3.60) in \cite{Dodelson:2003ft}), in this pre-thermal equilibrium WIMP scenario.

In a nutshell, the novelty of the pre-thermal equilibrium WIMP scenario is to include the phase of DM production. Although this phase is irrelevant to computing the relic abundance, it is crucial for investigating the evolution of DM density perturbation as we will show. Moreover, the above analysis on the background DM abundance has shown that, only during DM production ($y_{R_f}\le y\le y_\chi^{eq}$)\footnote{Since the process of reheating ($t_{R_i}\le t\le t_{R_f}$) is short and DM production is inefficient within this process, we have ignored such extra metric fluctuations during the reheating.}, the background is out of thermal equilibrium and has effective DM productions, so that only during this phase, the extra metric fluctuations, which are sourced by chemical potential difference between $\phi$ and $\chi$, can accumulate and affect the evolution of DM density perturbation. But during the thermal equilibrium phase ($y_\chi^{eq}\le y\le y_{fo}$) and the freezing-out phase ($y\ge y_{fo}$), no new extra metric fluctuations can not be produced. As we will see in next section, this observation can significantly simplify our analysis on the co-evolution of DM density perturbation and metric perturbation.

\section{Perturbative Resonance during Dark Matter production} \label{sec:Amplification}
In this section, we investigate the co-evolution of DM density perturbation and scalar modes of metric perturbation in the pre-thermal equilibrium WIMP scenario. As aforementioned, by decomposing the unintegrated Boltzmann equation in the early Universe up to the perturbative order, we can obtain not only Eq.(\ref{eq:eomback}) but also an improved Einstein-Boltzmann equation, which governs the co-evolution of DM density perturbation ($\delta\rho_\chi=-\rho_\chi\Theta$) and scalars mode of metric perturbation ({\it c.f.} Eq.(2.1) in \cite{Li:2019cjp}),
\begin{equation}\label{eq:eompert}
\frac{d\Theta}{dy}-3\frac{d\Phi}{dy}=\frac{\widetilde{\langle\sigma v\rangle}}{Hyn_\chi}\left[\left(\frac{n_\phi}{n_\phi^{eq}}\right)^2\left(n_\chi^{eq}\right)^2-n_\chi^2\right](\Theta+\Phi)~,
\end{equation}
where $\Theta(\vec{x},t)$ is the fluctuation of distribution function, $f=\exp[(\nu-E)/T][1-\Theta(\vec{x},t)]$, induced by the metric perturbation, $\Phi(\vec{x},t)$ and $\Psi(\vec{x},t)$ (see the definition of $g_{\mu\nu}$ at Eq.(\ref{eq:eomback})). Note that, for simplicity, we, in this work, only focus on the long wavelength (super-horizon) modes, leaving the short wavelength (sub-horizon) modes for further studies. $\Phi$, $\Psi$ and $\Theta$ are, respectively, the long wavelength Fourier modes of $\Phi(\vec{x},t)$, $\Psi(\vec{x},t)$ and $\Theta(\vec{x},t)$. For them, the terms with $\partial/\partial x_i$ have been neglected and the relation $\Phi=-\Psi$ is used.

To include the phase of DM production in the pre-thermal equilibrium WIMP scenario,  Eq.(\ref{eq:eompert}) is derived by discarding the thermal equilibrium assumption ($n_\chi=n_\chi^{eq}$ and $n_\phi=n_\phi^{eq}$). Therefore, this improved Einstein-Boltzmann equation contains an additional chemical-potential-difference-sourced term, $\left[\left(n_\phi/n_\phi^{eq}\right)^2\left(n_\chi^{eq}\right)^2-n_\chi^2\right]$, on the righthand side (RHS).  Comparing it with the RHS term of Eq.(\ref{eq:eomback}), we can immediately realize that this new term is just the perturbative part of the DM production term, which serves as an additional source for DM density at the perturbative order. Moreover, from the view of dynamical equation, this new term serves as a driving force that can affect the co-evolution of $\Theta$ and $\Phi$ during DM production. More specifically, it can amplifies them resonantly during this phase as we will show. Before solving the equation, we wish to point out that this new chemical-potential-difference-sourced term and its consequent effect do not involve the energy transportation between different spacetime locations  ($\partial/\partial x_i$ or $k_i$). Therefore, the resonant amplification of $\Theta$ and $\Phi$ (the super-horizon modes of perturbations) is a kind of local effect located at each spacetime location and does not violate the cosmological causality. To cross-check it, we can resubstitute the thermal equilibrium condition ($n_\chi=n_\chi^{eq}$ and $n_\phi=n_\phi^{eq}$) into Eq.(\ref{eq:eompert}), and it then reproduces the standard Einstein-Boltzmann equation,  $d\Theta/dy-3d\Phi/dy=0$ ({\it c.f.} Eq.(6.1) in Ref.\cite{Dodelson:2003ft}). And the amplitudes of these super-horizon modes, $\Theta$ and $\Phi$, become frozen again as expected in the conventional WIMP paradigm. 

To solve Eq.(\ref{eq:eompert}), we also need the perturbed Einstein equation in the radiation-dominated era ({\it c.f.} Eq.(6.6) in Ref.\cite{Dodelson:2003ft} and Eq.(2.3) in Ref.\cite{Li:2019cjp}), 
\begin{equation}\label{eq:perturbedeeq}
\frac{1}{H}\frac{d\Phi}{dt}+\Phi=-\frac{1}{2}\Theta~,
\end{equation}
where we have used the observation that all ultra-relativistic particles share the same $\Theta$ as it is induced by the same $\Phi$. By substituting Eq.(\ref{eq:perturbedeeq}) into Eq.(\ref{eq:eompert}), we obtain ({\it c.f.} Eq.(4.1) in Ref.\cite{Li:2019cjp}),
\begin{equation}\label{eq:secondorderphi}
\frac{d^2\Phi}{dy^2}+\frac{7}{2y}\frac{d\Phi}{dy}=\frac{\widetilde{\langle \sigma v \rangle}}{ H y n_\chi}\left[\left(\frac{n_\phi}{n_\phi^{eq}}\right)^2\left(n_\chi^{eq}\right)^2-n_\chi^2\right]\left(\frac{d\Phi}{dy}+\frac{1}{2y}\Phi\right)~.
\end{equation}
By solving it, we can obtain the whole evolution of $\Phi$ (and $\Theta$) as shown at following. 

In the pre-thermal equilibrium WIMP scenario, we can solve Eq.(\ref{eq:secondorderphi}) seperately in the three distinctive epochs, {\it Production} ($y_{R_f}\le y\le y_\chi^{eq}$), {\it Equilibrium} ($y_\chi^{eq}\le y\le y_{fo}$) and {\it Freezing-out} ($y\ge y_{fo}\simeq 1$). For the {\it Equilibrium} phase, the RHS term of Eq.(\ref{eq:secondorderphi}) is vanished as $n_\chi=n_\chi^{eq}$ and $n_\phi=n_\phi^{eq}$ in the thermal equilibrium. And for the {\it Freezing-out} phase, the RHS term of Eq.(\ref{eq:secondorderphi}) also vanishes as $(\widetilde{\langle \sigma v \rangle}n_\chi)/(H y)\rightarrow 0$. Therefore, during these two phases, the amplitude of $\Phi(y)$ is fixed and same to each initial value. And we only need to solve Eq.(\ref{eq:secondorderphi}) during the {\it Production} phase ($y_{R_f}\le y\le y_\chi^{eq}$).    

By substituting Eq.(\ref{eq:wholenwimp}) into Eq.(\ref{eq:secondorderphi}) and solving it with the initial conditions $\Phi(y_{R_f})=\Phi_\varphi$ and $d\Phi(y)/dy|_{y=y_{R_f}}=0$, we obtain
\begin{equation}\label{eq:phevi}
\Phi(y)=\Phi_\varphi \mathcal{G}\left[-y\kappa Y_{R_f}^{-1}\right], \quad y_{R_f}\le y\le y_\chi^{eq}, 
\end{equation}
where $\mathcal{G}(x)\equiv {_2F_1}\left(\frac{3-\sqrt{17}}{4},\frac{3+\sqrt{17}}{4};\frac{7}{2}; x \right)$ is short for the Gauss hypergeometric function\footnote{Mathematically, Gauss hypergeometric function, ${_2F_1}\left(a, b; c; x\right)$, is a solution of the hypergeometric differential equation, $x(1-x)y^{\prime\prime}+[c-(a+b+1)x]y\prime-aby=0$. And the decaying mode of the complete solution of Eq.(\ref{eq:secondorderphi}) is sub-dominated, so that it has been ignored in Eq.(\ref{eq:phevi}). }. This equation is same to the first line of Eq.(8) in Ref.\cite{Li:2019cjp} but we express the coefficient of the variable $y$ as $\kappa Y_{R_f}$ in this work.

According to Eq.(\ref{eq:phevi}), $\Phi(y)$ is amplified to be $\Phi(y_\chi^{eq})=\Phi_\varphi \mathcal{G}\left[-\left(2\pi^2Y_{R_f}\right)^{-1} \right]$ at the end of DM production ($y=y_\chi^{eq}$). And after that, $\Phi(y)$ becomes fixed,
\begin{equation}\label{eq:phevf}
\Phi(y\ge y_\chi^{eq})=\Phi_\varphi \mathcal{G}\left[-\left(2\pi^2Y_{R_f}\right)^{-1} \right],
\end{equation} 
since no DM particle is effectively produced at the Equilibrium and Freezing-out phase. Notably, according to this equation, the final value of $\Phi(y)$ is not only determined by its initial value $\Phi_{\varphi}$ but also relies on the value of $Y_f$, which is capsulated with the nature of DM particle and the physics of cosmic reheating.  

\begin{figure}[tbp]
\centering
\includegraphics[width=.8\textwidth]{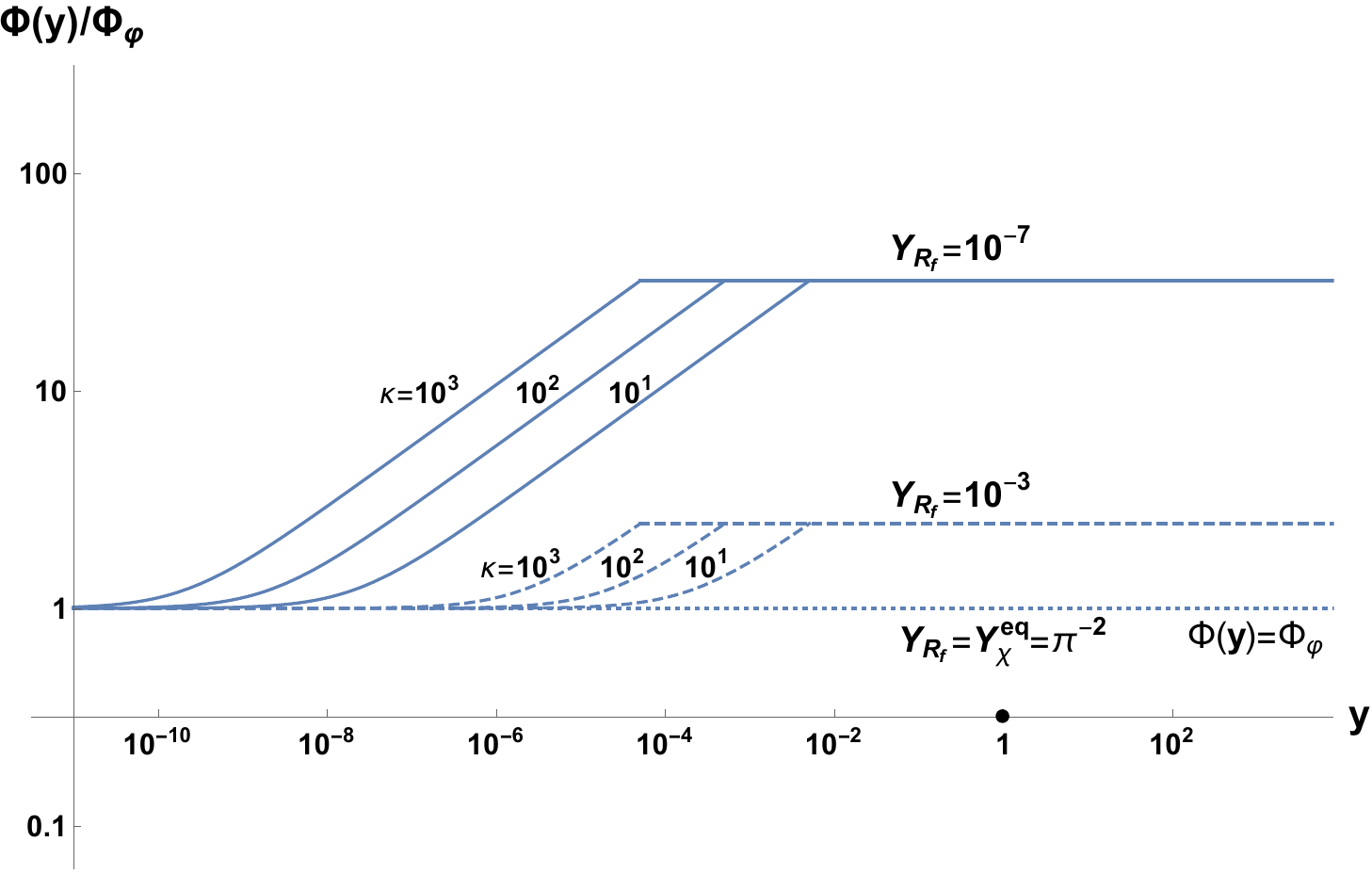}
\caption{\label{fig:Phi.pdf} The evolution of $\Phi(y)$ in the pre-thermal equilibrium WIMP scenario. After reheating, driven by the pair productions of DM particles, $\Phi(y)$ is amplified by the perturbative resonance. At the end of DM production ($y=y_\chi^{eq}$), the perturbative resonance ceases and the amplitude of $\Phi(y)$ becomes fixed. For the each case with the same value of $\kappa$, the amplification of $\Phi(y)$ ceases at the same time. And for each case with the same value of $Y_{R_f}$, the amplification of $\Phi(y)$ has the same magnitude. More specifically, this plot shows that a smaller value of $Y_{R_f}$ corresponds to a larger amplification of $\Phi(y)$. This is because a smaller value of $Y_{R_f}$ implies that a larger portion of DM abundance are produced after reheating, which can lead to a stronger perturbative resonance. In particular, for the case that DM particles are produced instantaneously during reheating ($Y_{R_f}\simeq Y_\chi^{eq}=\pi^{-2}$), we have no amplification of $\Phi(y)$ ($\Phi(y)=\Phi_\varphi$) and our solution becomes degenerate to the conventional WIMP paradigm (the horizontal dotted line). }
\end{figure}

In FIG.\ref{fig:Phi.pdf}, by using Eq.(\ref{eq:phevi}) and Eq.(\ref{eq:phevf}), we plot the evolution $\Phi(y)$ during the post-reheating epoch ($10^{-11}\le y\le 10^{4}$) with $Y_{R_f}=\{10^{-3}, 10^{-7}\}$ and $\kappa=\{10^1, 10^2, 10^3\}$. It clearly indicates that, driven by the pair productions of DM particles, $\Phi(y)$ is amplified significantly during DM production, and the amplification ceases when DM abundance attains thermal equilibrium. 

More specifically, FIG.\ref{fig:Phi.pdf} shows that, for the same value of $\kappa$, the amplification of $\Phi(y)$ ceases at the same time\footnote{Although this feature is not the target of this study, we want to emphasize that this feature may shed light on alleviating the well-known small-scale crisis as it implies a cut-off of the amplitude of $\Phi(y)$ at the small scale. In principle, by using more precise observation of sub-galactic structures, the value of $\kappa$ can be determined by such cut-off, so that $m_\chi$ can be constrained accordingly. We will leave this issue for further studies. And one can see Ref.\cite{Li:2019yyx} for a similar analysis in the NEQDM model. } as DM abundance attains  thermal equilibrium at $y_\chi^{eq}=(2\pi^2\kappa)^{-1}$. Moreover, we observe that for the same value of $Y_{R_f}$, the amplification of $\Phi(y)$ is the same. And a smaller value of $Y_{R_f}$ corresponds to a larger amplification of $\Phi(y)$. The reason is that a smaller value of $Y_{R_f}$ implies that a lager portion of DM abundance is produced after reheating. For the same initial metric inhomogeneity at the end of reheating, $\Phi(y_{R_f})=\Phi_\varphi$, a lager portion of DM abundance produced after reheating thus leads to a stronger amplification of $\Phi(y)$. Therefore, the final value of $\Phi(y)$ is inverse to the value of $Y_{R_f}$ as shown in FIG.\ref{fig:Phi.pdf}. To cross-check it, we can consider the case that all DM particles are produced instantaneously during reheating ($Y_{R_f}\simeq Y_\chi^{eq}=\pi^{-2}$) and no DM particles are produced after that. In this case, there is no amplification of $\Phi(y)$ and our solution is degenerate to the conventional WIMP paradigm ($\Phi(y)=\Phi_\varphi$, the horizontal dotted line). In a nutshell, this plot illustrates that the value of $Y_{R_f}$ can be traced from the final value of $\Phi(y)$. As aforementioned, $Y_{R_f}$ not only encodes the nature of DM particles but also contains the information of cosmic reheating. In principle, we can use $\Phi(y)$ to extract the particle nature of DM such as $m_\chi$. But we should notice that the particle nature of DM candidate and the parameters of reheating model are degenerate in $Y_{R_f}$ and in the amplification of $\Phi(y)$. Therefore, to trace $m_\chi$ from $\Phi(y)$, we need to specify the reheating process. For this task, we leave it in Section \ref{sec:ANewR}. Now we turn our attention to the cosmological implications on the amplification of $\Phi(y)$. 

As the E-modes in CMB has been precisely measured, one may worry about that the perturbative resonance may ruin the prediction of simple inflation \cite{Dodelson:2003ft}. But we can find that such perturbative resonance can only enlarge the amplitude of $\Phi$ but does not change the scale-dependence of its spectrum. Therefore, the perturbative resonance can preserve the consistence between the prediction of simple inflation and the observation of E-modes in CMB as the amplitude of primordial tensor modes (primordial B-modes) has not been determined observationally. Specifically, according to FIG.\ref{fig:Phi.pdf} (and Eq.(\ref{eq:phevf})), the amplification of $\Phi(y)$ is independent on the wavevector $k$, which indicates that, for each case with a fixed value of $Y_{R_f}$, the amplification for all long wavelength modes are the same. Thus, the scale-dependence of $\Phi(y)$ is preserved by the resonance, $d\ln P_{\Phi(y)}/d\ln k=d\ln P_{\Phi_\varphi}/d\ln k$, where $P_{\Phi(y)}$ and $P_{\Phi_\varphi}$ are the power spectra of $\Phi(y)$ and $\Phi_\varphi$ respectively. In particular, for the fiducial simple inflation, its scale-invariance is preserved, 
\begin{equation}\label{eq:pressi}
\frac{d\ln P_{\Phi(y)}}{d\ln k}=\frac{d\ln P_{\Phi_\varphi}}{d\ln k}\simeq 0~,
\end{equation}
which is consistent with current observations of E-modes in CMB \cite{Komatsu:2010fb, Ade:2015xua}.

Now we end this section by discussing the evolution of DM density perturbation, $\delta\rho_\chi=-\rho_\chi\Theta(y)$. By using Eq.(\ref{eq:perturbedeeq}), we obtain 
\begin{equation}\label{eq:thetaphi}
\frac{\Theta(y_\chi^{eq})}{\Theta(y_{R_f})}=\left.\left(\frac{y}{\Phi_\varphi}\frac{d\Phi(y)}{dy}+\frac{\Phi(y)}{\Phi_\varphi}\right)\right|_{y=y_\chi^{eq}}=\frac{\Phi(y_\chi^{eq})}{\Phi_\varphi}=\mathcal{G}\left[-\left(2\pi^2Y_{R_f}\right)^{-1} \right]~,
\end{equation}
where $d\Phi(y)/dy|_{y=y_\chi^{eq}}=0$ is used to derive the second step. This result indicates that, as $\Phi(y)$ and $\Theta(y)$ evolve resonantly, their amplifications are the same at the end of DM production. More importantly, this feature implies that the perturbative resonance can also preserve the consistence relation between the (dark) matter density distribution and the E-modes in CMB \cite{Li:2019cjp, Frenk:2012ph}. To see this, we can use Eq.(\ref{eq:thetaphi}) to obtain 
\begin{equation}\label{eq:thetaphifinal}
\delta\rho_\chi/\rho_\chi|_{y\ge y_\chi^{eq}}=\mathcal{G}\left[-\left(2\pi^2Y_{R_f}\right)^{-1} \right]=2\Phi(y)|_{y\ge y_\chi^{eq}},
\end{equation}
which is same to the standard characteristic relation obtained in the conventional WIMP paradigm, $(\delta\rho_\chi/\rho_\chi)_{standard}=2\Phi_\varphi=2\Phi_{standard}$, except both $\delta\rho_\chi/\rho_\chi$ and $\Phi$ are amplified in our model. Therefore, with Eq.(\ref{eq:pressi}) and Eq.(\ref{eq:thetaphifinal}), we can conclude that this chemical-potential-difference-sourced perturbative resonance can preserve the consistence amongst the simple inflation, the spectra of E-modes in CMB and the (dark) matter density distribution. But this conclusion also implies that, to constrain the perturbative resonance observationally, we need to go beyond the observations of E-modes and DM density distribution. In next section, we will introduce a strategy to constrain such perturbative resonance based on future observations of B-modes in CMB.

\section{The suppression of Tensor-to-scalar ratio of Metric Peturbation} \label{sec:TheSuppression}
In this section, we use the tensor-to-scalar ratio of metric perturbation to constrain the perturbative resonance. Since the perturbative resonance does not affect the tensor modes\footnote{By deriving the equation of motion for the tensor modes of metric perturbation, we can obtain $d^2h/d\eta^2+2aHdh/d\eta+k^2\eta=0$, which implies that the chemical potential difference does not affect the tensor modes.}, we can translate the amplification of $\Phi(y)$ into a suppression of the tensor-to-scalar ratio,
\begin{equation}\label{eq:revolvddmp}
r\equiv \frac{\mathcal{P}_h}{\mathcal{P}_{\Phi(y)}}=\frac{\mathcal{P}_h}{\mathcal{P}_{\Phi_\varphi}}\times\left(\frac{\Phi_\varphi}{\Phi(y)}\right)^2=r_i\times\left(\frac{\Phi_\varphi}{\Phi(y)}\right)^2=9\epsilon\times\left(\frac{\Phi_\varphi}{\Phi(y)}\right)^2~,
\end{equation}
where  the primordial power spectra of simple inflation $\mathcal{P}_{\Phi_\varphi}=\frac{8\pi G}{9 k^3}\frac{H^2_\star}{\epsilon}$ and $\mathcal{P}_h=\frac{8\pi G}{k^3}H^2_\star$, have been employed \cite{Dodelson:2003ft}, the slow-roll parameter takes $\epsilon\equiv dH^{-1}/dt\simeq 0.01$ \cite{Komatsu:2010fb, Ade:2015xua}, and $r_i=9\epsilon$ is the primordial value of $r$ at the end of reheating, which can be viewed as the result obtained by ignoring the process of DM production -- as aforementioned. 

According to Eq.(\ref{eq:revolvddmp}), when $\Phi(y)$ is amplified during DM production, $r$ is suppressed from its primordial value $r_i$. Once DM abundance attains thermal equilibrium at $y=y^{eq}$, such suppression ceases and $r$ attains its final value,  
\begin{equation}\label{eq:rzepi}
r=9\epsilon \left\{\mathcal{G}\left[-\left(2\pi^2\right)^{-1} Y_{R_f}^{-1}\right]\right\}^{-2},
\end{equation}
which is solely determined by the value of $Y_{R_f}$. 

\begin{figure}[tbp]
\centering
\includegraphics[width=.8\textwidth]{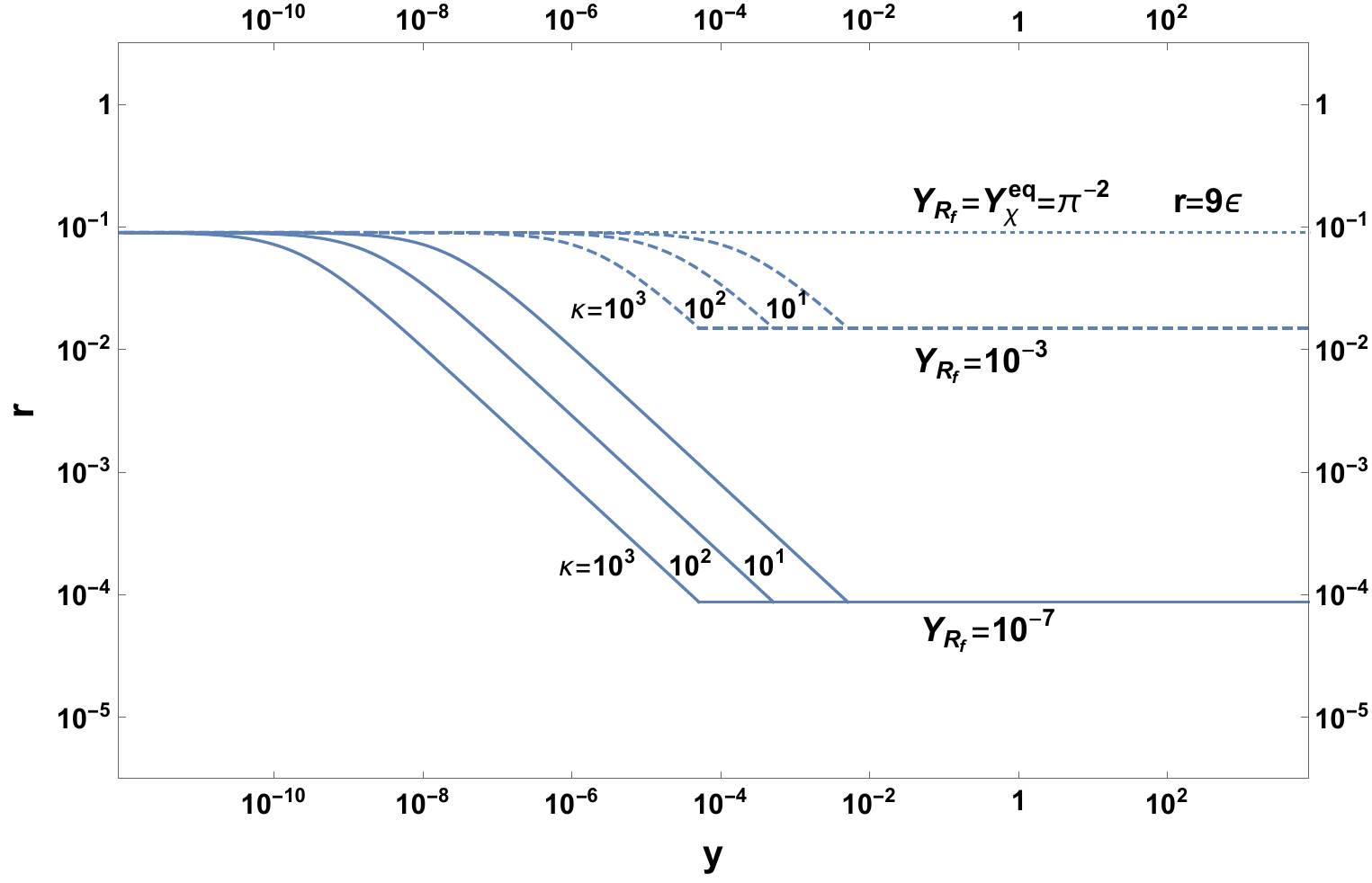}
\caption{\label{fig:r.pdf} The suppression of $r$ during the DM production phase in the pre-thermal equilibrium WIMP scenario. As the tensor modes of metric perturbation are not affected by the perturbative resonance, the amplification of $\Phi(y)$ results in a suppression of $r$ during the process of DM production. Loosely speaking, this plot is a reflection of FIG.\ref{fig:Phi.pdf}. }
\end{figure}

In FIG.\ref{fig:r.pdf}, by using Eq.(\ref{eq:revolvddmp}) and Eq.(\ref{eq:rzepi}), we plot the evolution $r$ during the post-reheating DM production epoch ($10^{-11}\le y\le 10^{4}$) with $Y_{R_f}=\{10^{-3}, 10^{-7}\}$ and $\kappa=\{10^1, 10^2, 10^3\}$. As expected, $r$ is suppressed during DM production and becomes fixed when DM abundance attains thermal equilibrium. More specifically, a smaller value of $Y_{R_f}$ corresponds to a stronger suppression of $r$ as it corresponds to a larger amplification of $\Phi(y)$ as shown in FIG.\ref{fig:Phi.pdf}. Again, for the case that all DM particles are produced instantaneously during the reheating ($Y_{R_f}\simeq Y_\chi^{eq}=\pi^{-2}$), the value of $r$ is degenerate to the well-known prediction of the conventional WIMP paradigm ($r\rightarrow r_i=9\epsilon$, the horizontal dotted line). 

In a nutshell, Eq.(\ref{eq:rzepi}) and FIG.\ref{fig:r.pdf} collectively indicate that, the perturbative resonance can generically result in a suppression of $r$ in comparing with the conventional prediction $r_i=9\epsilon$, which is based on the instantaneous DM production assumption. Therefore, once a suppression of $r$ is measured in coming round of PGW detections, it would imply that the perturbative resonance plays an important role during the production of pre-thermal equilibrium WIMPs. More explicitly, by specifying the value of $Y_{R_f}$ with a typical reheating model, we can use  Eq.(\ref{eq:rzepi})  to establish a new relation between $r$ and $m_\chi$ with two additional reheating parameters, $T_{R_f}$ and $\Gamma_0$, in the pre-thermal equilibrium WIMP scenario -- as presented in next section. In principle, using such relation, the particle nature of DM candidate and the reheating process can be constrained by future observations of $r$ in CMB.


\section{A new relation amongst Dark Matter, Cosmic Reheating and Primordial Gravitational Wave} \label{sec:ANewR}
Now we derive the relation between DM and PGW by specifying $Y_{R_f}$ in a simple reheating model \cite{Bassett:2005xm, Allahverdi:2010xz, Amin:2014eta}. Using  $\sigma_0=4\widetilde{\langle\sigma v\rangle}y^{-2}$ and $n_\phi=[T(t)]^{3}/\pi^{2}$ to simplify Eq.(\ref{eq:yrfexi}), we have 
\begin{equation}\label{eq:ychiexp}
Y_{R_f}=\frac{\sigma_0 m_\chi^2 }{4\pi^4 T_{R_f}^3}\int_{t_{m_\chi}}^{t_{R_f}}[T(t)]^4dt~, \quad t_{m_\chi}\le t\le t_{R_f}~,
\end{equation} 
where $T(t)$ characterizes how the Universe is reheated, and $t_{m_\chi}$ denotes the moment that $T=m_\chi$ during the reheating. As DM production is not efficient during $0<T\le m_\chi$, the integral for $t_{R_i}\le t\le t_{m_\chi}$ is neglected in this equation.  

By assuming that the inflaton $\varphi$ decays into SM particles effectively with a constant dissipative rate $\Gamma_0$, $\Gamma_0\gg 3(1+w_\varphi)H$, the equation of motion of $\rho_\varphi$ takes\footnote{In so-called ``old'' reheating theory, by assuming inflaton $\varphi$ decaying as a single-body, a rough estimation can relate $T_{R_f}$ and $\Gamma_0$, $T_{R_f}\simeq\sqrt{\Gamma_0 M_p}$. Thus a single parameter, $T_{R_f}$ or $\Gamma_0$, is enough in such theory (see Ref.\cite{Bassett:2005xm} for a review). However, in a general framework of cosmic reheating, the concept of single-body decay can not be always applied, so this work takes $T_{R_f}$ and $\Gamma_0$ as two independent parameters. } 
\begin{equation}\label{eq:rvpeq}
\frac{d\rho_\varphi}{dt}=-\Gamma_0\rho_\varphi.
\end{equation}
By substituting Eq.(\ref{eq:rvpeq}) into the continuity condition of energy density, 

$\sum_i\left[\frac{d\rho_i}{dt}+3(1+w_i)\rho_i\right]=0$, we obtain
\begin{equation}\label{eq:rgeq}
\frac{d\rho_\gamma}{dt}+4H\rho_\gamma=\Gamma_0\rho_\varphi,
\end{equation}
where $\sum_i$ denotes the summation of all relativistic components $\gamma$ and inflaton $\varphi$, and $w_i$ denotes the Equation of State for each component. On the other side, we have 
\begin{equation}\label{eq:rgt}
\rho_\gamma=g_\ast [T(t)]^4,
\end{equation}
where $g_\ast\simeq 90$ is the relativistic degrees of freedom at $t\ge t_{m_\chi}$. By solving Eq.(\ref{eq:rvpeq}), Eq.(\ref{eq:rgeq}) and Eq.(\ref{eq:rgt}), and substituting their solutions into Eq.(\ref{eq:ychiexp}) \footnote{ By solving Eq.(\ref{eq:rvpeq}), we obtain $\rho_\varphi=\rho_0 e^{-\Gamma_0 t}$, where $\rho_0$ is the value of $\rho_\varphi$ at $t=t_{R_i}=0$. By substituting it into Eq.(\ref{eq:rgeq}), we have 
\begin{equation}\label{eq:rhovex}
\frac{d\rho_\gamma}{dt}+4H\rho_\gamma=\Gamma_0\rho_0 e^{-\Gamma_0 t}~.
\end{equation}
For $t\le t_{R_f}$, as $\Gamma_0\gg H$, Eq.(\ref{eq:rhovex}) can be simplified as $d\rho_\gamma/dt=\Gamma_0\rho_0 e^{-\Gamma_0 t}$. And its solution is $\rho_\gamma=\rho_0 \left(1-e^{-\Gamma_0 t}\right)$. Then by substituting it into Eq.(\ref{eq:rgt}) and Eq.(\ref{eq:ychiexp}), we obtain 
\begin{equation}\label{eq:yrfint}
Y_{R_f}=\frac{\sigma_0 m_\chi^2 T_{R_f}}{4\pi^4}t_{R_f},
\end{equation}
where $\Gamma_0 t_{R_f}\gg 1$ and $t_{R_f}\gg t_{m_\chi}$ are used. By applying $d\rho_\gamma/dt |_{t=t_{R_f}}=0$ to Eq.(\ref{eq:rhovex}), we have $4H(t_{R_f})\rho_\gamma(t_{R_f})=\Gamma_0\rho_0e^{-\Gamma_0 t_{R_f}}$, which yields $t_{R_f}=-\Gamma_0^{-1}\ln\left[\frac{4\sqrt{2g_\ast}T_{R_f}^2}{M_p\Gamma_0}\right]$, where $H(t_{R_f})=\sqrt{2g_\ast}T_{R_f}^2/M_p$ are used. Then by substituting $t_{R_f}$ into Eq.(\ref{eq:yrfint}), we eventually obtain Eq.(\ref{eq:yrffi}).}, we obtain
\begin{equation}\label{eq:yrffi}
Y_{R_f}=-\frac{\sigma_0 m_\chi^2 T_{R_f}}{4\pi^4\Gamma_0}\ln\left(\frac{4\sqrt{2g_\ast}T_{R_f}^2}{M_p\Gamma_0}\right),
\end{equation}
where $M_p$ is the reduced Planck mass. Then, by substituting Eq.(\ref{eq:yrffi}) into Eq.(\ref{eq:rzepi}), we obtain an explicit relation amongst PGW, DM  and Reheating,
\begin{equation}\label{eq:rzep}
r=9\epsilon \left\{\mathcal{G}\left[\left(\frac{\sigma_0 m_\chi^2 T_{R_f}}{2\pi^2\Gamma_0}\ln\left[\frac{4\sqrt{2g_\ast}T_{R_f}^2}{M_p\Gamma_0}\right]\right)^{-1}\right]\right\}^{-2}.
\end{equation}

\begin{figure}[tbp]
\centering
\includegraphics[width=.8\textwidth]{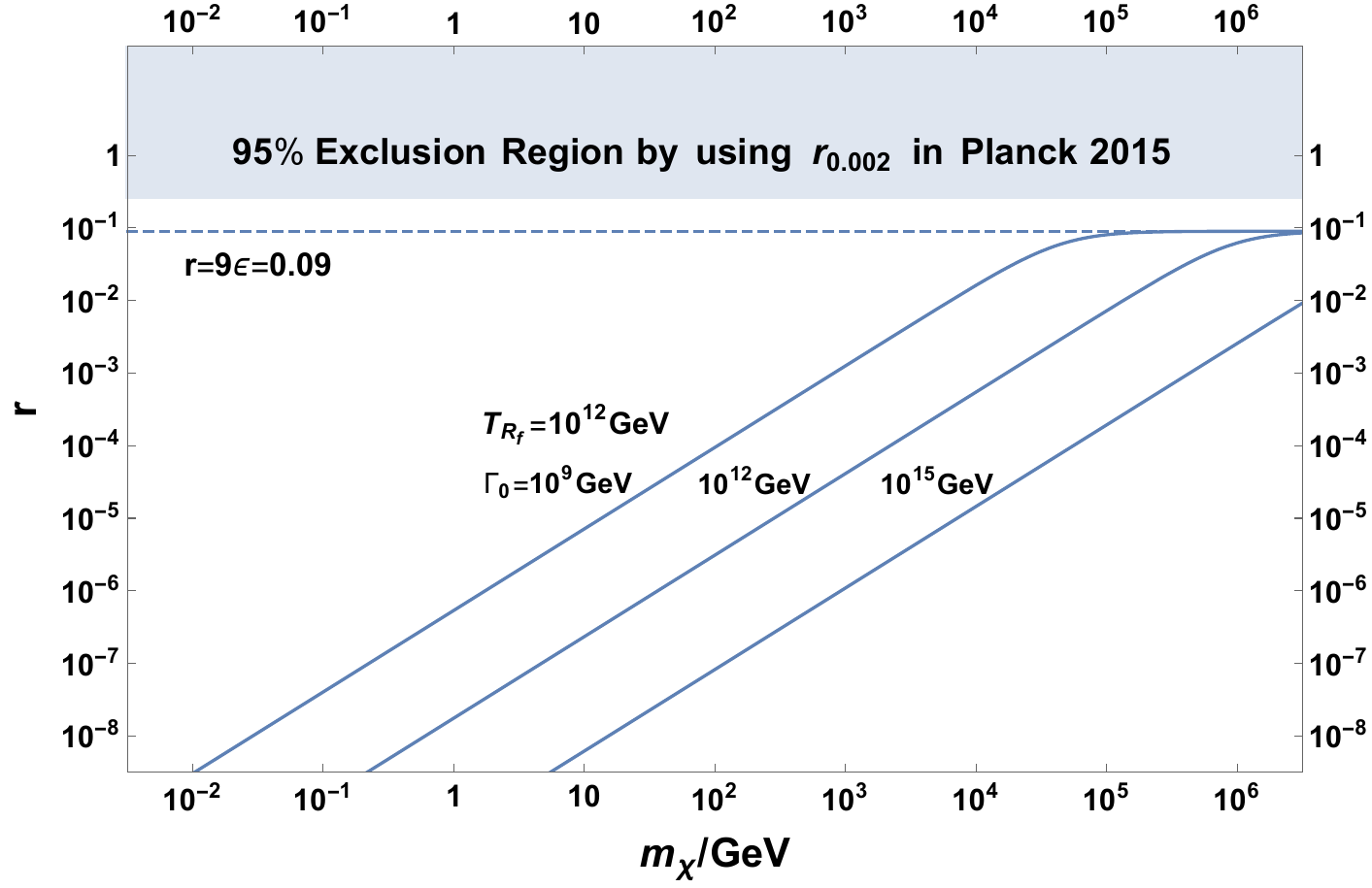}
\caption{\label{fig:rmchi.pdf} An illustration of the new relation between $m_\chi$ and $r$ with  $T_{R_f}=10^{12} \textbf{GeV}$ and $\Gamma_0=\{10^9,10^{12},10^{15}\} \textbf{GeV}$ in the pre-thermal equilibrium WIMP scenario. It shows that, for a given reheating process, a smaller value of $m_\chi$ corresponds to a smaller value of $r$ as it implies more DM particles produced after reheating, which drives a stronger perturbative resonance. On other hand, for a given $m_\chi$ and a fixed value of $T_{R_f}$, a larger value of $\Gamma_0$, which implies a shorter reheating process, corresponds to a smaller value of $r$. The reason is that a shorter reheating process implies a larger portion of DM abundance is produced after reheating. For the same initial metric perturbation at the end of reheating, $\Phi(y_{R_f})=\Phi_\varphi$, it leads to a stronger perturbation resonance and results in a smaller value of $r$. Reversely, for a larger $m_\chi$ and/or a smaller $\Gamma_0$, the suppression of $r$ is negligible and our solution is degenerate to the conventional WIMP paradigm, see the top right conner. And the shaded region on the top is the current observational limit on $r$ from Planck 2015 \cite{Ade:2015xua}.  }
\end{figure}

In FIG.\ref{fig:rmchi.pdf}, we illustrate Eq.(\ref{eq:rzep}) with $T_{R_f}=10^{12} \textbf{GeV}$ and $\Gamma_0=\{10^9,10^{12},10^{15}\} \textbf{GeV}$. We see that, for a sizeable parameter space, the value of $r$ is much smaller than the prediction of the conventional WIMP paradigm, $r\ll r_i=9\epsilon$. More specifically, for a given reheating process, a smaller value of $m_\chi$ corresponds to a stronger suppression of $r$. The reason is that, for a smaller value of $m_\chi$, more DM particles are produced after reheating and then results in a stronger amplification of $\Phi(y)$ and a heavier suppression of $r$. On the other hand, by fixing $m_\chi$ and $T_{R_f}$, a larger value of $\Gamma_0$ corresponds to a smaller value of $r$. The reason is that a larger value of $\Gamma_0$ implies a shorter reheating process, so that a larger portion of DM abundance is produced after reheating. For the same initial amplitudes of $\Phi(y)$ at the end of reheating, $\Phi(y_{R_f})=\Phi_\varphi$, a larger portion of DM abundance produced after reheating can thus lead to a larger amplification of $\Phi(y)$ during post-reheating DM production and then results in a smaller value of $r$ as shown in FIG.\ref{fig:rmchi.pdf}. Reversely, for a large $m_\chi$ and/or a small $\Gamma_0$ , the suppression of $r$ is negligible and our solution is degenerate to the conventional WIMP paradigm ($r\rightarrow r_i=9\epsilon$, see the top right conner). In a nutshell, once the value of $r$ is measured in future observations of B-modes in CMB, this relation and plot can be used to constrain $m_\chi$, $T_{R_f}$ and $\Gamma_0$ in principle. For example, supposing a reheating process with $T_{R_f}=10^{12}\textbf{GeV}$ and $\Gamma_0=10^9\textbf{GeV}$, if PGW is detected at $r=0.6\times 10^{-3}$ in future, $m_\chi$ is then predicted to be around $10^3\textbf{GeV}$. In FIG.\ref{fig:rmchi.pdf}, we additionally plot the current observational limit on $r$ based on Planck observation (the shaded region on the top) \cite{Ade:2015xua}. Although both our prediction (the solid curves) and the prediction of conventional WIMP paradigm (the dashed line) are still beyond the detectability of current experiments, we can expect that they can be reached and distinguished in the future \cite{Abazajian:2016yjj}.    

\section{Conclusion and Discussion} \label{sec:Conclusions}
In this paper, we investigate the co-evolution of DM density perturbation and scalar modes of metric perturbation in a newly introducing pre-thermal equilibrium WIMP scenario. Instead of assuming that DM particles are produced instantly, we embed a simple phase of DM production into the conventional WIMP paradigm, and extend our analysis to this out-of-chemical equilibrium phase. We find that, driven by the chemical potential difference between $\chi$ and $\phi$, a local perturbative resonance between DM density perturbation and the scalar modes of metric perturbation can also take place during DM production in the pre-thermal equilibrium WIMP paradigm and amplifies them resonantly. As such resonance does not affect the tensor modes, the amplification of scalar modes thus result in a suppression of tensor-to-scalar ratio of metric perturbation $r$, which predicts a smaller amplitude of PGW in comparing with the conventional prediction obtained by assuming WIMP DM starts in thermal equilibrium.

By analytically solving the improved Einstein-Boltzmann equation and using a typical reheating model to specify the cosmic background, we establish a new relation for pre-thermal equilibrium WIMP between DM particle mass $m_\chi$ and the tensor-to-scalar ratio $r$, which also contains two reheating parameters $T_{R_f}$ and $\Gamma_0$. Although the value of $r$ predicted by this relation is not yet reached by current observations of B-modes in CMB, we can expect that, optimized by the coming round PGW detections \cite{Abazajian:2016yjj}, this relation would be tested in the near future. Once a suppression of $r$ is measured in future, it would imply that the perturbative resonance plays an important role during the production of pre-thermal equilibrium WIMPs, and allow $m_\chi$, $T_{R_f}$ and $\Gamma_0$ to be constrained accordingly in principle.

Note that WIMP and NEQDM have different cosmic evolution (NEQDM never obtains thermal equilibrium); accordingly, the relation obtained for WIMP in this paper (Eq.(\ref{eq:rzep})) is different from that obtained for NEQDM \cite{Li:2019cjp}. Therefore, the discrepancy between these two relations also can be used to distinguish WIMP and NEQDM in principle, which is worthy of further study.

Moreover, we wish to emphasize that, as our proposal is based on DM density perturbation, it is alternative/complimentary to other strategies based on the background DM abundance. Therefore, a further study incorporating the typical constraints for WIMPs (e.g., direct: XENON1T; indirect: Fermi-LAT; collider: LHC) with the constraint obtained in this paper should be an exciting topic worthy of effort.    

To sum up, we highlight two issues for further studies. 
\begin{enumerate}
\item {\it The energy scale of inflation.} In this paper, we have demonstrated that, in the pre-thermal equilibrium WIMP scenario, the perturbative resonance during DM production can result in a smaller amplitude of PGW to be detected. It implies that, by taking into account DM production, the energy scale of inflation would be much smaller than the conventional expectation. In other words, if the perturbative resonance can be included, some inflation models, which have been excluded by their large tensor-to-scalar ratios, will be revitalized, such as the model with potential $V(\phi)\propto \phi^2$ \cite{Abazajian:2013vfg}. 
  
\item {\it On the small-scale structures.} As shown in FIG.\ref{fig:Phi.pdf}, the amplification of $\Phi$ completes at $y=y^{eq}$. For the long wavelength modes discussed in this work, the amplification is the same for all of them. However, for short wavelength modes, they get less amplification as they re-enter horizon before the end of DM production. It thus leads to a cut-off on the linear matter perturbation at small scale. By using horizon-crossing condition, the cut-off scale, $k_\star^{-1}$, can be determined as a function of $y^{eq}(m_\chi)$. If $k_\star^{-1}$ can be precisely measured in future \cite{Weinberg:2013aya}, $m_\chi$ can be constrained accordingly. Although the cut-off scale in the pre-thermal equilibrium WIMP scenario is smaller than that in NEQDM model \cite{Li:2019yyx}, it may also contribute to the formation of small-scale structures.

\end{enumerate}

At the end, we hope the perturbative resonance can be further explored in a broader scenario \footnote{In the two/multiple fields preheating/reheating models such as Refs.~\cite{Taruya:1997iv, Bassett:1998wg, Finelli:2000ya}, a resonant amplification of fields is also predicted, although there is no Boltzmann equation to govern the dynamics of fields. Therefore, it should be very interesting to explore how to coordinate these models and the current work within a broader landscape in the future. } with various DM candidates \cite{Tanabashi:2018oca} and various alternative cosmological models \cite{Nojiri:2017ncd}, and it may help us better understand the physics at the earliest epoch of our Universe.

\section{Acknowledgments}

We thank Yeuk-kwan Edna Cheung, Xiaheng Xie and Yajun Wei for useful discussion, careful reading of this manuscript and important suggestion on improving the clarity of presentation. We appreciate the anonymous referee for helping us to clarify many important issues in this paper. This work has been supported by the National Natural Science Foundation of China ( 11603018, 11963005, 11775110, 11433004, 11690030 ) and Yunnan Provincial Grants ( 2016FD006, 2019FY003005, 2015HA022, 2015HA030 ).

\section*{References}

\bibliography{mybibfile}

\end{document}